\documentclass[12pt]{article}
\usepackage{amsmath, amssymb}
\usepackage{amsfonts}
\usepackage[pdftex]{graphicx}

\usepackage[hmargin=2cm,vmargin=2cm]{geometry}

\usepackage{setspace}
\begin{document}

\begin{center}
\textbf{Towards a Quantitative Description of Solid Electrolyte Conductance Switches}
\end{center}

\begin{center}
Monica Morales-Mas\'{\i}s$^*$, Hans-Dieter Wiemh$\ddot{\text{o}}$fer, Jan M. van Ruitenbeek
\end{center}

[$^*$]   MSc. M. Morales-Mas\'{\i}s, Prof.dr. J. M. van Ruitenbeek\\
Kamerlingh Onnes Laboratorium\\
Universiteit Leiden, PO Box 9504, 2300 RA Leiden, The Netherlands \\
E-mail: morales@physics.leidenuniv.nl

~~~~~Prof.dr. H.D. Wiemh$\ddot{\text{o}}$fer\\
Institut f$\ddot{\text{u}}$r Anorganische und Analytische Chemie\\
Universit$\ddot{\text{a}}$t M$\ddot{\text{u}}$nster, Corrensstr. 30, 48149 M$\ddot{\text{u}}$nster, Germany\\

[$^{**}$]
\textbf{Acknowledgments}

We thank Sense Jan van der Molen for numerous helpful discussions, and Federica Galli for AFM technical support. This research is carried out with financial support from the Dutch Foundation for Fundamental Research on Matter (FOM).\\


\newpage

\begin{abstract}

We present a quantitative analysis of the steady state electronic transport in a resistive switching device. The device is composed of a thin film of Ag$_{2}$S (solid electrolyte) contacted by a Pt nano-contact acting as ion-blocking electrode, and a large area Ag reference electrode. When applying a bias voltage both ionic and electronic transport occurs, and depending on the polarity it causes an accumulation of ions around the nano-contact. At small applied voltages (pre-switching) we observed this as a strongly nonlinear current-voltage curve, which have been modeled using the Hebb-Wagner treatment for polarization of a mixed conductor. This model correctly describes the transport of the electrons within the polarized solid electrolyte in the steady state up until the resistance switching, covering the entire range of non-stoichiometries, and including the supersaturation range just before the deposition of elemental silver. In this way, it is a step towards a quantitative understanding of the processes that lead to resistance switching.  \\

\end{abstract}


\section{Introduction}

Resistance switching based memories, show the potential to be integrated into future micro-electronic components. Such devices, consisting of oxide materials or solid electrolytes, present novel properties that allow for scalability down to the nano and even atomic scale, and very low power consumption. \textsuperscript{\cite{Waser2007, Waser2009, Terabe1}}\\

With this increased interest in memory resistive devices, there is a necessity to understand the physical mechanisms driving the resistance switching process. Various models have been proposed to explain the switching mechanism, many of these models in agreement with the conductive filament formation, and annihilation, inside the insulator material.\textsuperscript{\cite{Waser2007, Jo2009, Liu2009}} Nevertheless, a deep understanding of the microscopic mechanism responsible for filament formation is still lacking.\\

In this paper we demonstrate the use of the Hebb-Wagner formalism \textsuperscript{\cite{Wagner1957, HEBB1952}} for the analysis of the steady state I-V characteristics of memory resistors based upon mixed ion and electron conductors. We apply this formalism to fit our experimental I-V characteristics, and describe the ionic and electronic transport within the electrolyte before the full resistance switching is observed.\\

The mixed conductor used for the present study is Ag$_{2}$S. However, this description could also be valid for other mixed electronic and ionic conductors, e.g. Cu$_{2}$S and AgGeSe. In general, one has to take into account that the formation of a space charge layer occurs for many solid electrolytes. In the case of Ag$_{2}$S, effects due to depletion or space charge at the Pt contact are negligible. The device we consider consists of a Ag$_{2}$S thin film contacted by a Ag thin film at the bottom (which helps to achieve a reference state with constant silver concentration at that contact), and a nano-scale Pt contact on the top realized by means of a conductive AFM tip.\\

Our measurements and simulations at low bias voltages (steady state) confirm the predictions of the theory: the increase in the electronic current at forward bias (negative polarity at the Pt contact) is due to the initial accumulation of Ag$^{+}$-ions towards the nano-scale contact. This causes a Ag concentration gradient, i.e. local deviations from the ideal stoichiometry in the region close to the nano-contact. We note that this occurs before reduction of the Ag$^{+}$-ions and therefore before any switching is observed.\\

The Hebb-Wagner concepts have originally been formulated for bulk materials, and until today, to our knowledge, have not been applied to nano-scale contacts or thin film devices. We demonstrate that the theory still holds for nano-scale devices. By confirming this, we can achieve a better understanding of the conditions of ionic and electronic transport in mixed conductors and solid electrolytes which lead to conductance switching.

\section{Results}

\begin{figure}[h]
  \begin{center}
    \includegraphics[width=7cm]{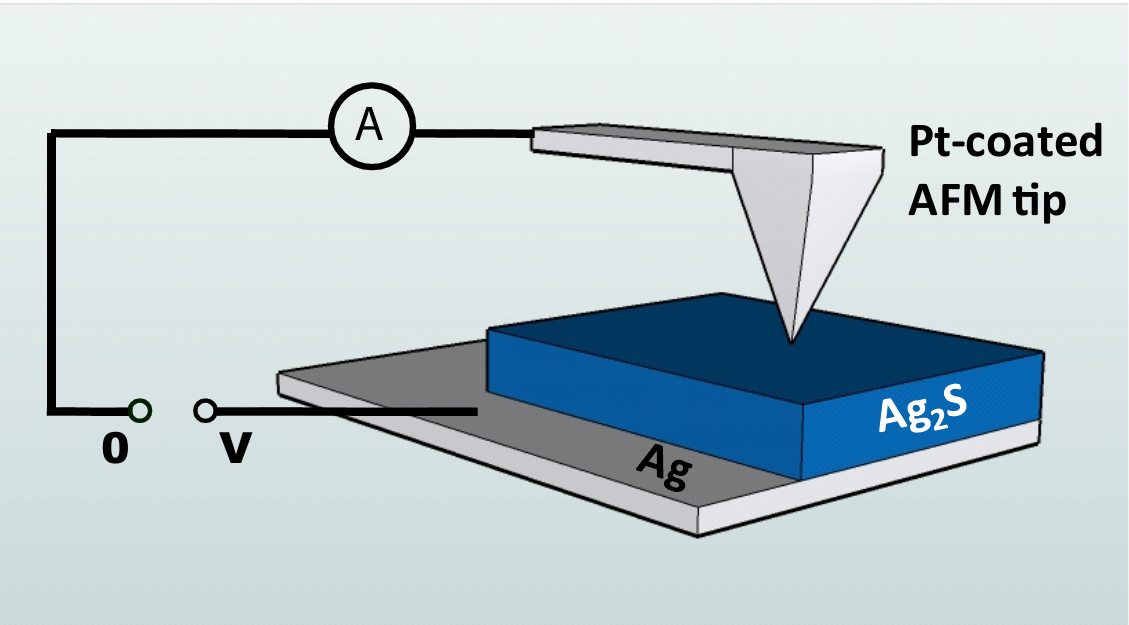}
        \caption{Schematic diagram of the electrode configuration for the measurements with the Conductive Atomic Force Microscope (CAFM)}
  \label{tip} 
  \end{center}
\end{figure}

The electrical measurements were performed with the use of a conductive atomic force microscope (C-AFM)  (Veeco Multimode AFM/SPM system). In the setup, the Ag layer is the bottom contact to the Ag$_{2}$S layer and the top contact is a Pt-coated AFM tip. See the diagram in Figure \ref{tip}.\\

Platinum is a chemically inert metal, and as an electrode in the system it blocks the ionic current. By using a nano-scale Pt electrode and a reference (Ag) electrode at the bottom with a large surface area, the changes in electrical conductivity will be concentrated at the vicinity of the nano-contact.\\ 

The current-voltage characteristics of the samples were obtained by continuously ramping the voltage linearly from 0 to $V_{max}$ down to $-V_{max}$ and back to 0, at a frequency of 0.25 Hz. Voltages are given throughout with respect to the potential of the Pt tip (taken as 0V). For the steady state analysis, the value of $V_{max}$ was kept below the potential at which we observed hysteresis in the I-V characteristics,\textsuperscript{\cite{Morales-Masis2009}} meaning that no significant changes are induced in the solid electrolyte by decomposition of the Ag$_{2}$S.\textsuperscript{\cite{lehmann}} All the experiments were performed at ambient conditions. \\

\begin{figure}[ht]
  \begin{center}
        \includegraphics[width=6cm]{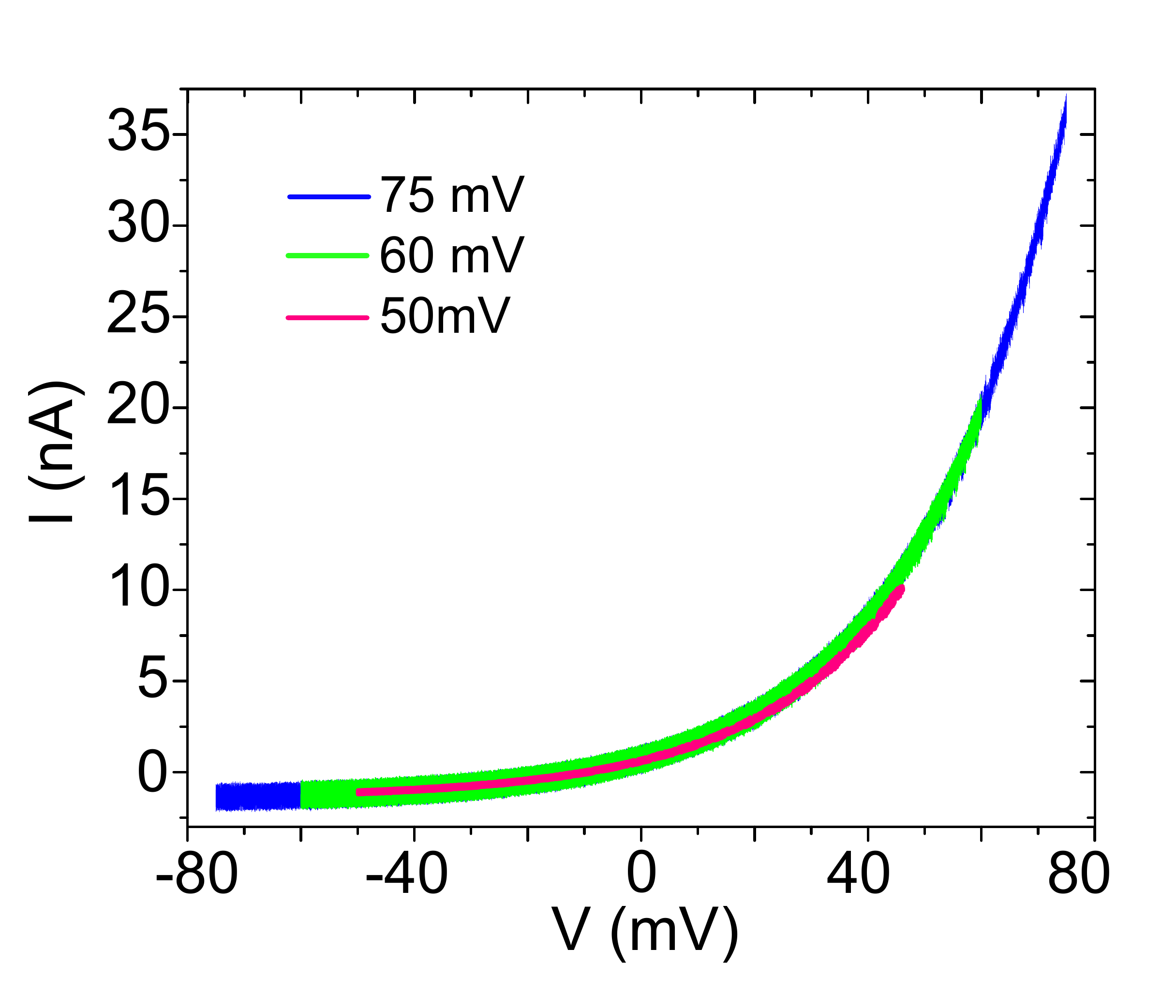} 
        \includegraphics[width=6cm]{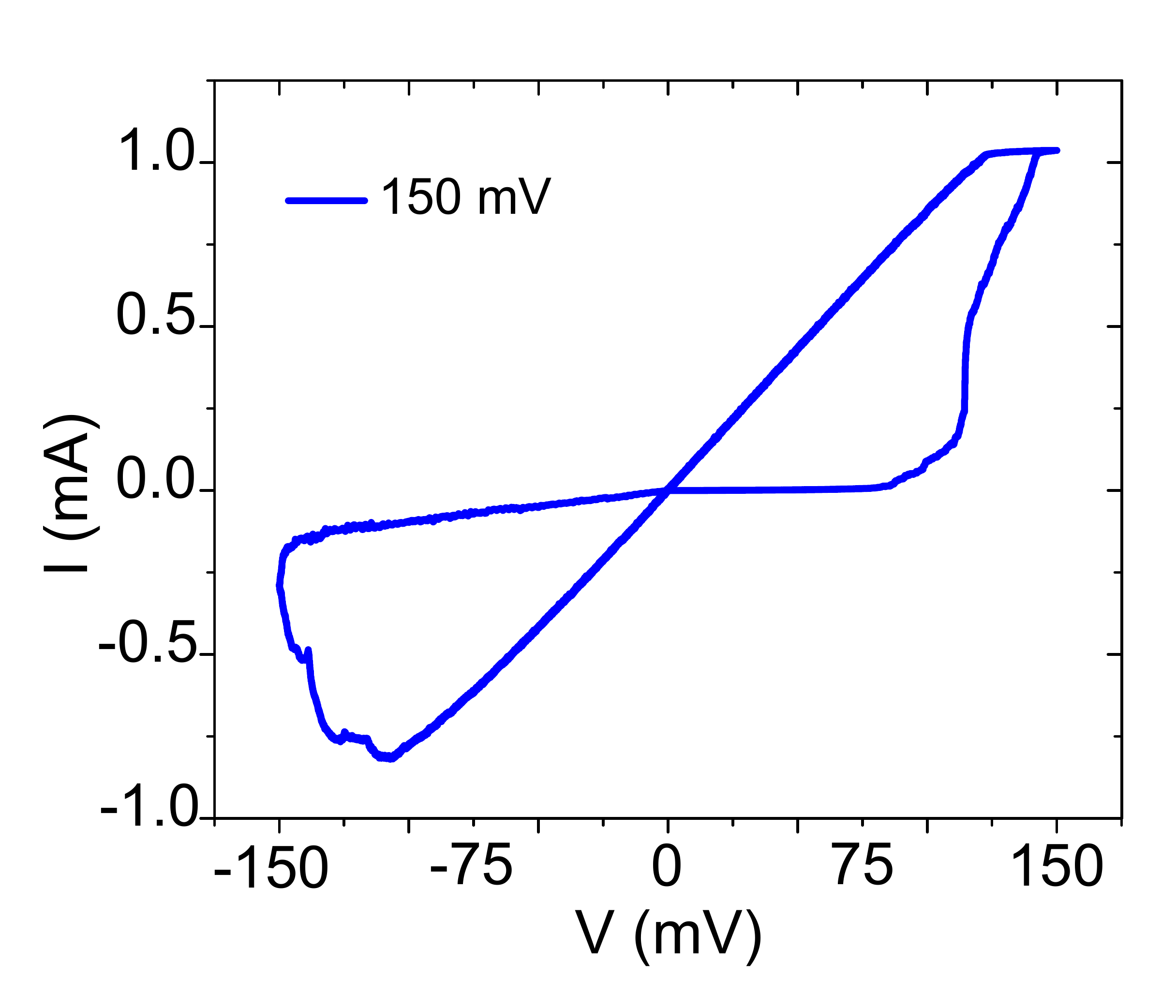}
     \caption{Steady State (left panel) and full bipolar switching (right panel) current-voltage characteristics of the Ag/Ag$_{2+\delta}$S/Pt(nano-contact) system}
    \label{IVs}
  \end{center}
\end{figure}

The current-voltage characteristics show an exponential behavior that is fully reversible on the time scale of the experiment. The curve is asymmetric, with an increase in the current at the positive bias. We refer to this as the 'pre-switching' steady state behavior and, in the Ag/Ag$_{2+\delta}$S/Pt junctions, it is observed for voltages below about 75mV (left panel Figure \ref{IVs}). When increasing the bias voltage beyond 75 mV, the I-V curves present hysteresis, which evolves into full bipolar switching for still larger voltages (right panel Figure \ref{IVs}). In the full bipolar switching case, the ON and OFF states of the device are clearly observed, with a resistance ratio (R$_{\text{OFF}}$/R$_{\text{ON}}$) of approximately 10$^{5}$.\\ 

 For the purpose of this paper, we focus on the exponential, or steady-state, I-V curves (left panel of Figure \ref{IVs}). The shape of the curve is reminiscent of curves measured for metal-semiconductor junctions, and it is known that Ag$_{2+\delta}$S is a n-type semiconductor.\textsuperscript{\cite{Kashida2003}} However, the obtained curves have the inverse curvature as compared to that expected for a Pt/n-type semiconductor junction (i.e. Schottky contact). In the case of solid electrolyte semiconductors, this must be attributed to a combination of ionic and electronic conduction in the electrolyte. The main observation is the fact that the shape of the IV-curves depends strongly on the type of electrode use: material, size and symmetry of the electrodes.\textsuperscript{\cite{Bard}} \\

 In our experiment, with the use of a nano-scale contact and a much larger bottom contact, we have introduced an asymmetry in the potential distribution across the mixed conductor. When a potential difference is applied to the system Ag/Ag$_{2+\delta}$S/Pt, where Pt is the nano-contact, the strength of the electric field is concentrated in the vicinity of the nano-contact. If the potential is small, such as to avoid decomposition of Ag$_{2+\delta}$S, a steady state composition gradient is induced in the Ag$_{2+\delta}$S film, as a result of the mobility of Ag$^+$-ions.\textsuperscript{\cite{HEBB1952}}\\ 

 For a mixed electronic and ionic conductor such as Ag$_{2+\delta}$S, the electronic conductivity is a function of the deviations from the stoichiometric composition ($\delta$). When the negative polarity is at the Pt tip (nano-contact), the Ag$^+$-ions move towards the tip, acting as n-type donors. The local enhancement of the Ag ion concentration results in an increase of the electronic conductivity in the small region close to the tip. We will elaborate on this below.\\ 

\subsection{Theory}

A model for the current-voltage behavior of mixed ionic conductors under steady state conditions goes back to Hebb and Wagner.\textsuperscript{\cite{Wagner1957, HEBB1952}}\\
 
Silver sulfide is a mixed conductor as both silver ions and electrons are mobile. Applying a voltage $V$ to a silver sulfide sample between two electrodes, sets up a difference of the Fermi levels, $\varepsilon^{''}_{\text{F}}$ and $\varepsilon^{'}_{\text{F}}$, between both contacts. \\

If we can neglect the interface resistances, the applied electrical potential difference $V$ imposes a difference of the local electrochemical potentials of electrons in the ionic conductor between the two electrode interfaces on the mixed conductor

\begin{equation}
		-e \: V = \varepsilon^{''}_{\text{F}} - \varepsilon^{'}_{\text{F}} = \tilde{\mu}^{''}_{e} - \tilde{\mu}^{'}_{e}   
\label{eq1}   
\end{equation}

where $-e$ is the electron charge. The values at the boundaries are denoted as, prime for the Ag bottom contact and double prime for the Pt nanocontact.\\

According to Eq.(\ref{eq1}), applying a voltage generates a gradient of the electrochemical potential $\tilde{\mu}_{e}$ within the mixed conducting silver compound and, thus, an electronic current density $j_{e}$ given by

\begin{equation}
\stackrel{\rightarrow}{j}_{e} = \frac{\sigma_{e}}{e} \stackrel{\rightarrow}{\nabla} \tilde{\mu}_{e} 
\label{Xje}   
\end{equation}

with $\sigma_{e}$ as the conductivity of electrons. \\   

A gradient of the electrochemical potential of electrons is accompanied by a gradient of the electrochemical potential of silver ions, and vice versa. The corresponding silver ion current density is given by a complementary expression to Eq.(\ref{Xje}) according to

\begin{equation}
\stackrel{\rightarrow}{j}_{\text{Ag}^{+}} = -\frac{\sigma_{\text{Ag}^{+}}}{e} \stackrel{\rightarrow}{\nabla} \tilde{\mu}_{\text{Ag}^{+}}
\label{jAg}	
\end{equation}

with $\sigma_{\text{Ag}^{+}}$ as the conductivity of silver ions. \\

Furthermore, the currents of electrons and silver ions will be coupled by the equilibrium of electrons and ions according to

\begin{equation}
		\text{Ag} \rightleftharpoons \text{Ag}^{+} + e^{-}  
\label{equil}   
\end{equation} 

The assumption of local thermodynamic equilibrium between silver ions and electrons is valid, if the electrochemical potential gradient in Eq.(\ref{Xje}) is not too high.\textsuperscript{\cite{Wagner1933}} This approach is well accepted for mixed conducting silver chalcogenides.\textsuperscript{\cite{Mizusaki1975}} Then, the condition of thermodynamic equilibrium holds for Eq.(\ref{equil}) at all positions in the sample. With $\mu_{\text{Ag}}$ denoting the chemical potential of neutral silver, it follows 

\begin{equation}
		\mu_{\text{Ag}} = \widetilde{\mu}_{\text{Ag}^{+}} + \widetilde{\mu}_{e}  
\label{chem-pot}   
\end{equation}  

According to the local equilibrium condition (\ref{chem-pot}), the boundary condition Eq.(\ref{eq1}), imposed by the voltage applied between the ion blocking Pt nano-contact and the silver back contact leads also to the following alternative expression for the applied voltage

\begin{equation}
 -e \: V = \left( \mu^{''}_{\text{Ag}} - \mu^{'}_{\text{Ag}} \right) - \left( \tilde{\mu}^{''}_{\text{Ag}^{+}} - \tilde{\mu}^{'}_{\text{Ag}^{+}} \right)
\label{eq6}	
\end{equation}
 
Let us consider two simple limiting cases which lead to simplified equations for the voltage. The first concerns a sample with a homogeneous composition initially. At $t=0$, just after a sudden jump of the voltage from $V = 0$ to $V > 0$, the chemical potential of neutral silver has a constant value throughout the whole sample. Under this condition, the difference of the chemical potentials of silver in Eq.(\ref{eq6}) vanishes and accordingly, the remaining difference of the electrochemical potentials of silver ions in Eq.(\ref{eq6})  will be identical to the difference of the electrochemical potentials of electrons in magnitude but opposite in sign.

\begin{equation}
t = 0:  \qquad -e \: V  =  \mu^{''}_{e} - \mu^{'}_{e} = - \left( \tilde{\mu}^{''}_{\text{Ag}^{+}} - \tilde{\mu}^{'}_{\text{Ag}^{+}} \right) 
\label{eq7a}	
\end{equation}

For increasing times $t > 0$, the initial silver ion current builds up a concentration gradient in the sample. This leads to an increasing value of the chemical potential difference $\left(\mu^{''}_{\text{Ag}} - \mu^{'}_{\text{Ag}}\right)$ and a decreasing electrochemical potential difference of silver ions which finally reaches zero. Accordingly, the electrochemical potential difference of silver ions, and hence the ionic current, will vanish for long enough times $t >> 0$ giving

\begin{equation}
t \gg 0:  \qquad -e\:  V  =  \mu^{''}_{e} - \mu^{'}_{e} =  \mu^{''}_{\text{Ag}} - \mu^{'}_{\text{Ag}} 
\label{eq7}	
\end{equation}

The metallic silver bottom electrode fixes the chemical potential at the interface Ag/Ag$_{2}$S at $\mu^{\circ}_{\text{Ag}}$. Because of this, the chemical potential of silver (and accordingly the non-stoichiometry $\delta$) at the ion-blocking electrode, is the only variable in the system which is linearly dependent on the applied voltage. Therefore, under steady state conditions, Eq.(\ref{eq7}) simplifies to

\begin{equation}
		-e \: V = \mu^{\circ}_{\text{Ag}} - \mu^{'}_{\text{Ag}}   
\label{eq10}   
\end{equation}

Note in this case that for $V=0$, the chemical potential of silver at the ion-blocking contact is equivalent to that of metallic silver. Therefore, if no supersaturation occurs, any positive voltage should lead to the formation of metallic silver deposits. However, this is not observed for positive voltages up to 75 mV meaning that around the ion-blocking contact, a supersaturated composition occurs with $\delta >\delta^\circ$ where $\delta^\circ$ corresponds to the non-stoichiometry in thermodynamic equilibrium with silver metal. \\

Hence, in the steady state, the gradient of the electrochemical potential of silver ions and the ionic current density vanish. Then, the gradient of the electrochemical potential of electrons is equivalent to the gradient of the chemical potential of silver, and the total electric current density $j_{\text{total}}$ is carried by the electrons only. This is summarized in the following,

\begin{align}
 t \gg 0:        \qquad  \nabla \mu_{\text{Ag}^{+}}  =    0 , 
 								 \qquad   \ \ \ \; j_{\text{Ag}^{+}} =    0 \\
                 \qquad  \nabla \mu_{e}              =    \nabla \mu_{\text{Ag}}, 
                 \qquad   j_{\text{total}}           =     j_{e} 
\label{eq11}
\end{align}

Therefore, in the steady state, the electronic current can be expressed by the gradient of the chemical potential of neutral silver according to

\begin{equation}
		t \gg 0: \qquad j_{e} = j_{\text{total}} = \frac{\sigma_{e}}{e} \nabla \mu_{\text{Ag}}
\label{eq12}
\end{equation}

The typical time to reach the steady state as assumed for Eqs.(\ref{eq10}) to (\ref{eq12}) is estimated as $\tau$ = $L^{2}_{\text{diff}}$/2 $D_{\text{Ag}}$, with $\tau$ denoting the relaxation time for building up a steady-state silver concentration gradient in the sample, $L_{\text{diff}}$ the diffusion length through the sample (for a linear geometry the distance between the two electrodes), and $D_{\text{Ag}}$ the chemical diffusion coefficient of silver that is given by \textsuperscript{\cite{RICKERT1983}}

\begin{equation}
D_{\text{Ag}} = \frac{\sigma_{e} \sigma_{\text{Ag}^{+}}}{\sigma_{e} + \sigma_{\text{Ag}^{+}}} \cdot \frac{1}{e^{2}} \cdot \frac{d\mu_{\text{Ag}}}{dc_{\text{Ag}}} 
\label{diff}
\end{equation}

with c$_{\text{Ag}}$ denoting the local concentration of Ag atoms  in Ag$_{2+\delta}$S. \\

All quantities on the right side of Eq.~(\ref{diff}) are available from the literature.\textsuperscript{\cite{Bonnecaze}} The chemical diffusion coefficient in the low temperature phase of Ag$_{2+\delta}$S is very high, at 80$^{\circ}$C the value is of the order of 10$^{-2}$ cm$^{2}$/s. At ambient conditions, it is still around 10$^{-5}$ cm$^{2}$/s. Therefore, decay of concentration gradients in Ag$_{2+\delta}$S occurs much faster than in many other mixed conducting solids. The reason is that the thermodynamic factor $\frac{d\mu_{\text{Ag}}}{dc_{\text{Ag}}}$ in Eq.(\ref{diff}) has an extremely large value due to the small range of non-stoichiometry.\textsuperscript{\cite{Bonnecaze}}\\

It has to be remarked that the relations discussed above are only valid as long as no deposition of metallic silver occurs. If silver metal is formed near and at the ion-blocking contact, the ion-blocking boundary condition is no more valid and a continuous silver ion current will flow. The total current then is a sum of electronic and ionic currents and will not reach a steady state. At a nano-contact, it will usually increase, because the formation of metallic silver will virtually increase the contact area of the nano-contact. Finally, one expects a short-circuiting of the electrodes by the grown silver filaments. \\

When the steady-state conditions with no silver deposition are fulfilled, Eq.(\ref{eq12}) can be applied and a well defined, unambiguous relation exists between the electronic conductivity and the chemical potential of silver. Following a concept by C. Wagner,\textsuperscript{\cite{Wagner1957}} one can calculate the electronic conductivity from the slope of the steady state I-V curve. First, the chemical potential and the space variable are separated according to

\begin{equation}
		e \; \stackrel{\rightarrow}{j}_{\text{total}}(\stackrel{\rightarrow}{r}) \cdot \: d\stackrel{\rightarrow}{r} = \sigma_{e} \: d\mu_{\text{Ag}}
\label{eq13}
\end{equation} 

The relation is given for a general case where the current density may vary along the coordinates $\stackrel{\rightarrow}{r}$. Integrating Eq.(\ref{eq13}), with integration limits set by the boundary conditions at the two electrode contacts, we have

\begin{equation}
		e \int^{r^{'}}_{r^{''}} \stackrel{\rightarrow}{j}_{\text{total}}(\stackrel{\rightarrow}{r}) \cdot \: d\stackrel{\rightarrow}{r} = \int^{\mu^{'}_{\text{Ag}}}_{\mu^{0}_{\text{Ag}}} \sigma_{e} \; d\mu_{\text{Ag}}
\label{eq14}
\end{equation} 

The left-hand side of Eq.(\ref{eq14}) depends on the geometry of the contacts. We will simplify the integrals by assuming the chemical potential drops only along one coordinate, say $r$. The integral becomes simple for a sharp point contact (radius a) and a hemispherical reference electrode at $r^{''}\rightarrow \infty$. In this case we have,

\begin{equation}
		e \int^{r^{'}}_{r^{''}} j_{\text{total}}(r) \;dr = -\frac{e}{2 \pi a} \: I 
\label{eq14b}
\end{equation} 
 
More generally we obtain an expression with the total electrical current I and a constant $K$ that depends on the distribution of the current density through the sample: 

\begin{equation}
		e \int^{r^{'}}_{r^{''}} \stackrel{\rightarrow}{j}_{\text{total}}(\stackrel{\rightarrow}{r}) \cdot \: d\stackrel{\rightarrow}{r} = -\frac{e}{K} \: I 
\label{eq15}
\end{equation} 

with K = A/L for a specimen in the shape of a pellet with area A and thickness L, K = 2$\pi$a for a hemispherical contact of radius a, and K = 4a for a disk shaped contact to a semi-infinite sample.\\
						
With this result for the left hand side of Eq.(\ref{eq14}), following C. Wagner \textsuperscript{\cite{Wagner1933}} one can differentiate both sides with respect to the upper integration boundary $\mu^{'}_{\text{Ag}}$. This yields
 
\begin{equation}
		-\frac{e}{K}\cdot\frac{dI}{d\mu^{'}_{\text{Ag}}} = \frac{d}{d\mu^{'}_{\text{Ag}}}\left( \int^{\mu^{'}_{\text{Ag}}}_{\mu^{0}_{\text{Ag}}} \sigma_{e} \; d\mu_{\text{Ag}} \right) = \sigma_{e}\left(\mu^{'}_{\text{Ag}}\right)
\label{eq16}
\end{equation} 

From Eq.(\ref{eq10}) we have $\mu^{'}_{\text{Ag}} = \mu^{0}_{\text{Ag}} + e \: V$, accordingly $d\mu^{'}_{\text{Ag}} = e \: dV$ and  

\begin{equation}
		\left(\frac{dI}{dV}\right)_{\text{steady \! state}} = - K \sigma_{e}(\mu^{'}_{\text{Ag}}) 
\label{eq17}
\end{equation} 

Now that we have eliminated the unknown chemical potential we can integrate again over the electrical potential to obtain the final current-voltage relation in the steady state 

\begin{equation}
I(V) = - \int^{0}_{V} K \cdot \sigma_{e}(V) dV
\label{eq18}	
\end{equation}
 
In our case, we are working in the range of stoichiometry $\delta$ $>$ 0, i.e. the n-type range. Under this assumption the electronic conductivity is given by $\sigma_{e} = \sigma_{0} e^{(eV/kT)}$.\textsuperscript{\cite{Mott}} The voltage dependence arises from the relation between the local Ag$^+$-ion concentration (doping) and the local electrical potential.\\

Thus, we obtain

\begin{equation}
I(V) = K \sigma_{0} \: \frac{k_{B}T}{e}  \left(e^{(eV/k_{B}T)} - 1\right) 
\label{I(V)}	
\end{equation}

where $k_{B}$ is Boltzmann's constant and $K$ is a constant representing the geometrical factor mentioned in Eq.(\ref{eq15}). In the limit near zero bias, i.e. eV $<<$ kT, Eq.(\ref{I(V)}) reduces to $I = K \sigma_{0} V$.\\ 

As indicated in Eq.(\ref{eq10}), the relation above applies only in the cases when one of the electrodes maintains a constant chemical potential for Ag, and all changes occur at the small electrode. In other experiments,\textsuperscript{\cite{Morales-Masis2009}} by using two Pt electrodes with the same asymmetry in the geometry, we observed a slight deviation from the theory due to additional chemical potential changes at the Pt bottom contact. From a thermodynamic point of view, in principle I-V relations could also be calculated, if one knows the initial non-stoichiometry $\delta$ of the sample. But as they depend on the initial $\delta$, it is not easy to achieve well defined experimental conditions.\\     

In order to approach the limit of the semi-infinite sample (no changes at the bottom electrode) we need to make use of a very small contact. This allow us also to test the theory for nanometer size limit.\\ 

\section{Discussion}

\begin{figure}[ht]
  \begin{center}
        \includegraphics[width=6cm]{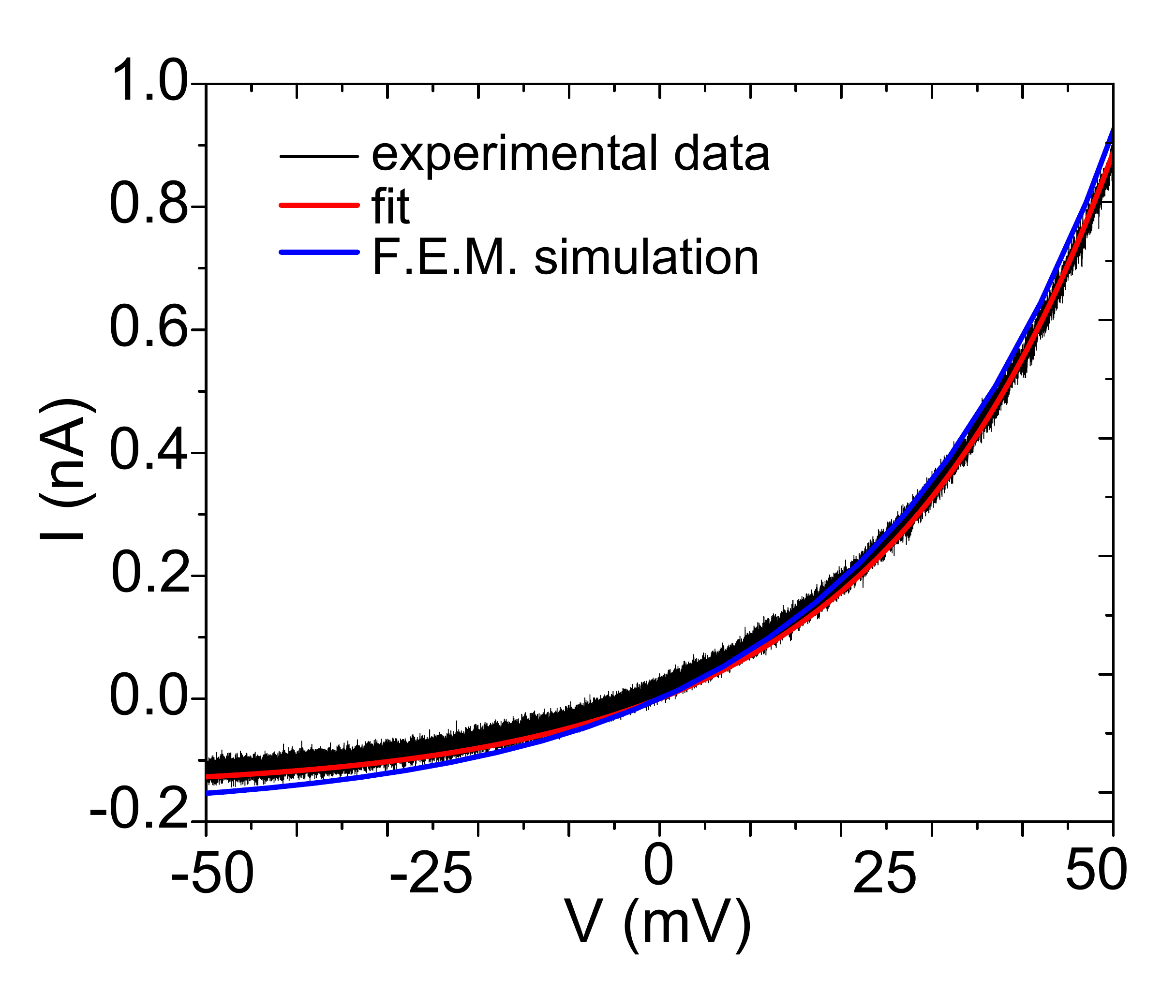} 
        \includegraphics[width=6cm]{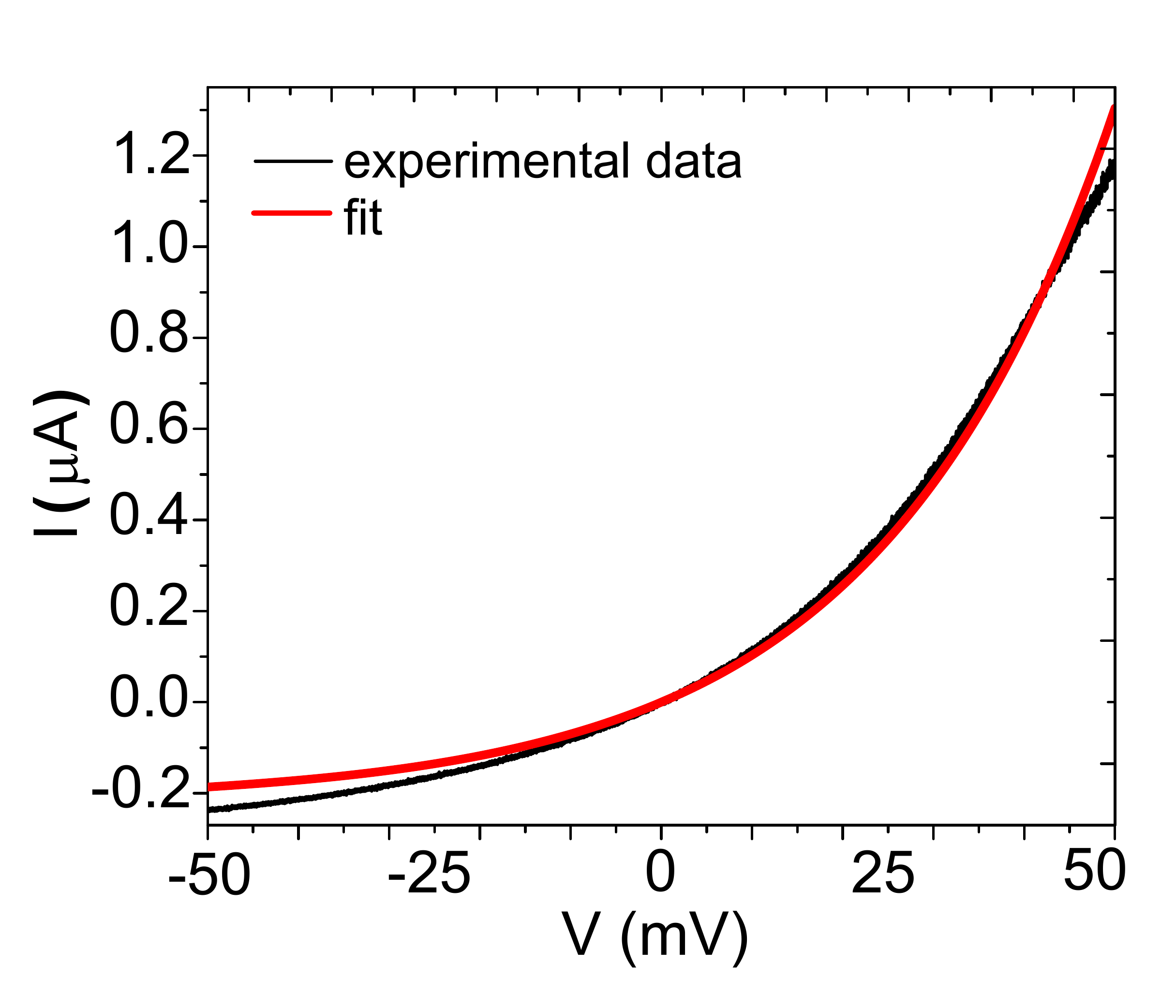}
     \caption{Steady state current-voltage characteristics of the Ag/Ag$_{2+\delta}$S/Pt system, measured with a Pt-coated AFM tip (left panel) and a Pt wire (right panel). Fit with Eq.(\ref{I(V)}) are shown as the red curves.}
    \label{IV-fit}
  \end{center} 
\end{figure}

Figure \ref{IV-fit} shows two sets of data for a 200nm thick Ag$_{2}$S film on top of a Ag bottom reference electrode (black lines). The curve at the left is taken with a Pt coated AFM tip at very small load in order to minimize the contact size. Systematic measurements of tip load with contact size were performed for calibration (See Supporting Information). The curve at the right is taken with a Pt wire contact for a larger macroscopic contact.\\

Using $\sigma_{0}$ = 7.8 x 10$^{-2}$ $\Omega^{-1}m^{-1}$ \textsuperscript{\cite{Bonnecaze}} and T = 298 K, we can compare the experimental data of the steady state I-V curves with Eq.(\ref{I(V)}). The result of the fitting is shown in figure \ref{IV-fit} (red lines).\\

\begin{figure}
\centering
\begin{tabular}{cc}
   \includegraphics[width=8cm , angle=270]{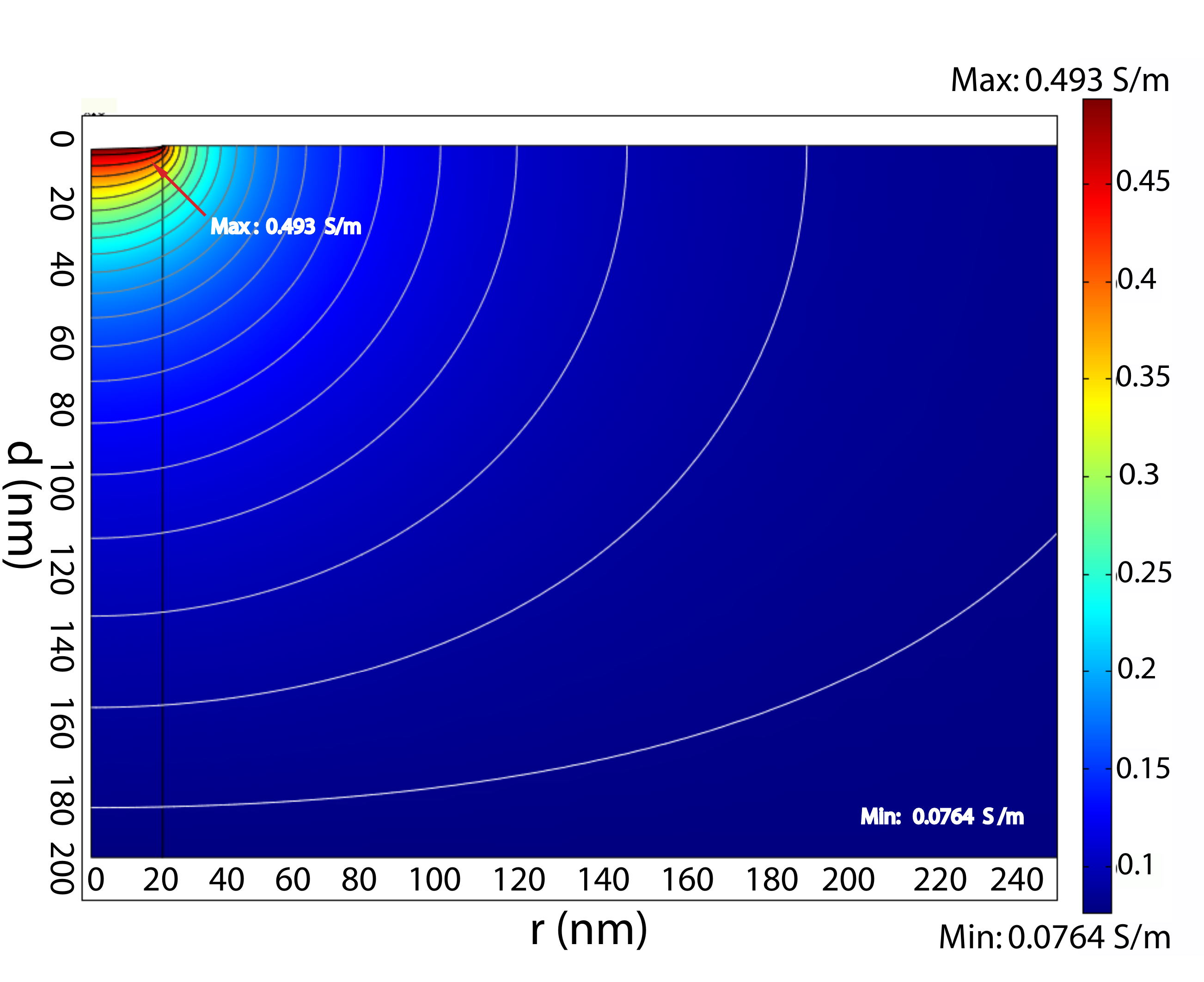} \\
   \includegraphics[width=8cm]{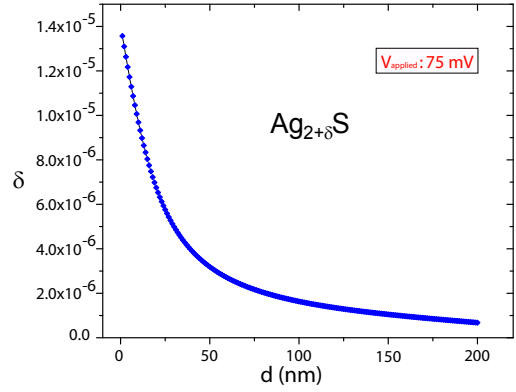}  \\
\end{tabular}
\caption{Top panel: Finite element simulation of potential (contour plot) and conductivity (surface plot) at V = 52mV, across the 200 nm thick Ag$_{2+ \delta}$S. Contact radius is 20nm, with a nearly planar geometry. Bottom panel: Plot of the non-stoichiometry $\delta$ as a function of the vertical distance from the center of the nano-contact to the Ag substrate (d). The values of $\delta$ indicate a strong increase in Ag$^+$-ion concentration at the region neighboring the nano-contact reaching the values of high supersaturation of Ag in Ag$_{2+ \delta}$S.}
\label{sigma_simu}
\end{figure}

The geometry of the electrodes and sample enters through the constant $K$ in Eq.(\ref{I(V)}), which is the only adjustable parameter. Assuming that the end of the AFM tip has approximately hemispherical shape, K = 2$\pi$a , we obtain an estimate for the tip contact radius, a, 

\begin{equation}
a = K / (2 \pi) = 12 \text{nm} \nonumber \\ 
\label{r}	
\end{equation} 

which agrees well with the AFM tip radius observed by electron microscopy of approximately 20 nm. Using finite element  simulations, we have modeled the sample and electrodes geometry (Figure \ref{sigma_simu}). The resulting total current as a function of the applied voltage (same as used for the measured IV's) results in a IV curve reproducing those obtained from the experiment and fitting.\\  

For comparison to the measurements with the AFM tip, we have also used a macroscopic Pt wire of 0.1 mm diameter as the top contact. The steady state IV-curve as well as the corresponding fitting with Eq.(\ref{I(V)}) is shown in the right side of Figure \ref{IV-fit}. The quality of the fit is much less good than in the case of the nano-contact. This, is because the assumption of a semi-infinite sample is not longer valid. In our case, fitting with Eq.(\ref{I(V)}) led to a calculated effective contact radius of 17 $\mu$m. However, assuming a perfect contact, the total current must be much higher than observed in Figure \ref{IV-fit}. In our experiment, the Pt wire will not be a perfect contact, but rather the total area in contact with the Ag$_2$S surface is reduced to only few contact points distributed in an effective radial area of $\approx$ 17 $\mu$m.\\  

For visualization of the top electrode size effect, we have simulated the voltage drop and local conductivity across the thickness of the Ag$_{2}$S sample. The model geometry is the same as used for the measurements: a Ag large bottom electrode, a 200nm thickness Ag$_{2}$S layer and a 20nm radius Pt top electrode with a nearly planar geometry. For the simulation we used axial symmetry around the nano-contact.\\

Figure \ref{sigma_simu} (top panel) shows the solution of the finite element simulation at a voltage of 52 mV (voltage polarity as defined for the experiment). The figure clearly shows that the conductivity changes are concentrated in the region close to the top electrode. Also, we observed that the conductivity increases when the negative polarity is at the top electrode, indicating a local increase of Ag$^+$-ion concentration (n-type donors). In the opposite polarity, the conductivity will decrease, due to the local depletion of Ag$^+$-ions, leading to a lower conductivity. The color bars at the right side of the simulation indicate the conductivity values ($\sigma_e$).\\  

The bottom panel of Figure \ref{sigma_simu} is a plot of the variation in stoichiometry obtained from the simulation results (in this case at a voltage of 75 mV). It is plotted as a function of the vertical distance (d) from the center of the nanocontact to the Ag substrate. The deviation from stoichiometry is given by,\textsuperscript{\cite{Bonnecaze, BECKER1983}}

\begin{equation}
		 \delta = n - p = 2 \: K_i^{1/2} \! \sinh \left(\frac{e(V_0-V(d))}{kT}\right)       
\label{delta}   
\end{equation} 

where n and p are the electron and hole concentration and $K_i^{1/2}$ = n = p at the stoichiometric composition. In the calculation we have taken $K_i^{1/2}$ = 2 x 10$^{20}$ m$^{-3}$ and $V_0$ = 105 mV as reported by Bonnecaze $\textit{et al.}$\textsuperscript{\cite{Bonnecaze}} The equation above accounts for both p an n-type Ag$_{2+ \delta}$S.\\

The range of homogeneity of the low temperature phase $\alpha$-Ag$_2$S is extremely narrow, with $\left|\delta^\circ\right|$ in the order of 10$^{-6}$.\textsuperscript{\cite{SCHMALZRIED1980, BECKER1983}} For $\delta > 0$, Ag$_2$S exist with excess Ag (n-type regime), and for $\delta < 0$, there is deficit of Ag (p-type regime). We define $\delta^{\circ}$ as the non-stoichiometry limit in thermodynamic equilibrium with Ag (at the n-type regime). In our experiment this is the case where the Ag$_2$S is in contact with the Ag substrate. \\

In Figure \ref{sigma_simu} we observe that $\delta$ = $\delta^{\circ}$ at the boundary Ag/Ag$_2$S (d = 200nm) as expected, and $\delta$ $>$ $\delta^{\circ}$ over the whole range of the curve. This would mean that Ag should precipitate from the Ag$_2$S already over the full range. However, a certain level of supersaturation is needed before Ag starts precipitating. When a certain supersaturation level is reached, precipitation of metallic Ag will begin. In our geometry we will therefore observe precipitation of metallic Ag to start in the region near the nano-contact (Pt tip). This precipitation process is, in a later stage, responsible for the nucleation and further formation of Ag filaments. The formation of these filaments, which can grow to make a metallic contact between the Ag and Pt electrode, is the proposed mechanism that leads to full bipolar conductance switching.\\

The nucleation process is related to an overpotential threshold, below which nucleation is basically zero and above which nucleation will increase exponentially.\textsuperscript{\cite{Waser2009, Budevski2000}} It is also remarkable that the nucleation of silver metal occurs within the silver sulfide phase and does not start from an adsorbed silver layer on the blocking electrode. This makes the resistance switching unsensitive to the nature of the blocking electrode. \\

The precipitation of metallic silver during our measurements from an oversaturated state can be understood from the $IV$ curves, explaining hysteresis and full conductance switching. We observed metal deposition and switching to high conduction at a bias voltage beyond 75~mV, a value in agreement with earlier results.\textsuperscript{\cite{Rickert871983}} As indicated above, the experimental data verify that there is a nucleation barrier for the formation of the metallic silver, that can be related with e.g. lattice deformation and surface free energy. However, further studies are needed to clarify the background for the observed threshold voltage for deposition of Ag and the complete description for the system beyond the critical supersaturation. \\

\section{Conclusion}

We present above a quantitative analysis of the steady state ionic and electronic transport in a solid electrolyte device that leads to resistance switching. The model presented here describes the electronic transport within the solid electrolyte in the steady state, covering the range of non-stoichiometries due to additional Ag in Ag$_2$S, up to the supersaturation range just before the deposition of elemental silver. The model is then a base for a complete description of solid electrolyte conductance switches, and it can be extended to other semiconductor materials with mobile donors or acceptors.\\ 

\section{Experimental Section}

The Ag$_{2}$S devices are fabricated as follows: first a layer of Ag (100nm thickness and 10 x 10 mm$^{2}$ surface area) is sputtered onto a Si(100) substrate covered with a native oxide layer. On top of the Ag layer, the Ag$_{2}$S layer (200nm thickness and 5 x 5 mm$^{2}$ surface area) is grown by sputtering of Ag in a Ar/H$_{2}$S plasma, with the use of a shadow mask. For the preparation of stoichiometric Ag$_{2}$S by RF-sputtering, the most important parameter is the partial pressure of H$_{2}$S in the sputtering atmosphere. To estimate the partial pressure of H$_{2}$S in the sputtering chamber, we measure the total pressure (P$_{\text{total}}$ = P$_{\text{H}_{2}\text{S}}$ + P$_{\text{Ar}}$) with a Compact Capacitance Gauge (CMR 264, Pfeiffer Vacuum). First the Ar partial pressure is set to establish the sputtering discharge (pre-sputtering), and before deposition on the substrate, H$_{2}$S is introduced in the chamber. To achieve a stoichiometric composition, the partial pressure of H$_{2}$S used is $\approx$ 6 x 10$^{-4}$ mbar.\\ 

\begin{figure}[ht]
  \begin{center}
    \includegraphics[width=7cm]{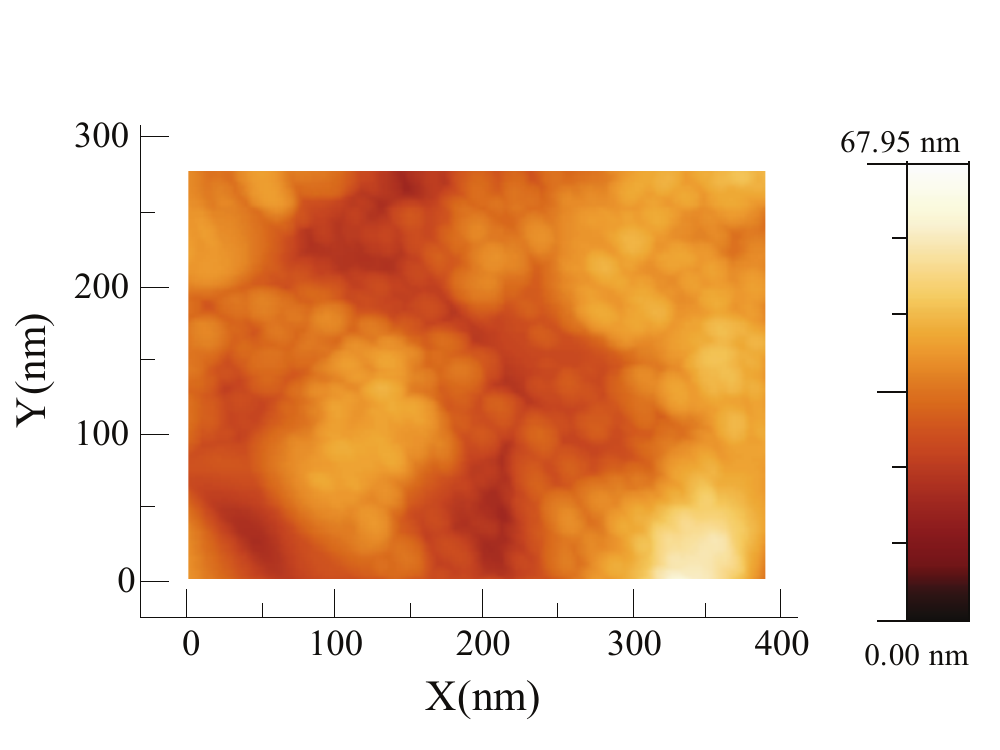}
        \caption{AFM topography image of the Ag$_{2}$S surface. The right color code shows the scale bar for the height in the image.}
  \label{surface} 
  \end{center}
\end{figure}

 Composition of the samples is analyzed by X-Ray Diffraction (XRD) and Energy Dispersive X-Ray analysis (EDX). The surface morphology of the as-prepared samples is also checked by Atomic Force Microscopy (AFM) and Scanning Electron Microscopy (SEM). Figure \ref{surface} shows an AFM image of the sample surface. We observe a rough surface with grain sizes of approximately 20 to 40 nm diameter.\\

\newpage


\begin{thebibliography}{10}

\bibitem{Waser2007}
R.~Waser, M.~Aono, \newblock {\em Nat. Mater.} \textbf{2007}, \textit{6}, 833-840. 

\bibitem{Waser2009}
R.~Waser, R.~Dittmann, G.~Staikov, K.~Szot, \newblock {\em Adv. Mater.} \textbf{2009}, \textit{21}, 2632-2663.

\bibitem{Terabe1}
K.~Terabe, T.~Hasegawa, T.~Nakayama, M.~Aono, \newblock {\em Nature}, \textbf{2005}, \textit{433}, 47-50.

\bibitem{Jo2009}
S.~H. Jo, K.~H. Kim, W.~Lu, \newblock {\em Nano Lett.}, \textbf{2009}, \textit{9}, 496-500.

\bibitem{Liu2009}
Q.~Liu, C.~M. Dou, Y.~Wang, S.~B. Long, W.~Wang, M.~Liu, M.~H. Zhang, J.~N.
  Chen, \newblock {\em Appl.Phys.Lett.}, \textbf{2009}, \textit{95}, 023501.

\bibitem{Wagner1957}
C.~Wagner, \newblock {\em Proc. 7th Meeting Intern. Committee on Electrochemical Thermodynamics and Kinetics, Lindau 1955}, \newblock {\em Butterworth Scientific Publ., London}, \textbf{1957}, 361.
  
\bibitem{HEBB1952}
M.~H. Hebb, \newblock {\em J. Chem. Phys.}, \textbf{1952}, \textit{20}, 185-190. 

\bibitem{Morales-Masis2009}
M.~Morales-Masis, S.~J. van~der Molen, W.~T. Fu, M.~B. Hesselberth, J.~M. van  Ruitenbeek, \newblock {\em Nanotechnology}, \textbf{2009}, \textit{20}, 095710.

\bibitem{lehmann}
V.~Lehmann, H.~Rickert, \newblock {\em J. Appl. Electrochem.}, \textbf{1979}, \textit{9}, 209-217.

\bibitem{Kashida2003}
S.~Kashida, N.~Watanabe, T.~Hasegawa, H.~Iida, M.~Mori, S.~Savrasov,
\newblock {\em Solid State Ionics}, \textbf{2003}, \textit{158}, 167-175.

\bibitem{Bard}
A.~J.~Bard, L.~R.~Faulkner, \newblock {\em Electrochemical Methods: Fundamentals and Applications / 2nd ed.}, \newblock John Wiley \& Sons, Inc., \textbf{2001}.

\bibitem{Wagner1933}
C.~Wagner, \newblock {\em Z.Physik.Chem.}, \textbf{1933}, \textit{25}, B21.

\bibitem{Mizusaki1975}
J.~Mizusaki, K.~Fueki, T.~Mukaibo, \newblock {\em Bull. Chem. Soc. Jpn.}, \textbf{1975}, \textit{48}, 428-431.

\bibitem{RICKERT1983}
H.~Rickert, H.~D.~Wiemhofer, \newblock {\em Solid State Ionics}, \textbf{1983}, \textit{11}, 257-268.

\bibitem{Bonnecaze}
U.~Bonnecaze, A.~Lichanot, S.~Gromb, 
\newblock {\em J. Phys. Chem. Solids}, \textbf{1978}, \textit{39}, 299-310.

\bibitem{Mott}
N.F. Mott, R.W. Gurney, \newblock {\em Electronic Processes in Ionic Crystals / 2nd ed.}, \newblock Oxford University Press, \textbf{1950}.

\bibitem{BECKER1983}
K.~D. Becker, H.~Schmalzried, V.~Vonwurmb,
\newblock {\em Solid State Ionics}, \textbf{1983}, \textit{11}, 213-219.

\bibitem{SCHMALZRIED1980}
H.~Schmalzried, \newblock {\em Prog. Solid State Chem.}, \textbf{1980}, \textit{13}, 119-157.

\bibitem{Budevski2000}
E.~Budevski, G.~Staikov, W.J.~Lorenz, \newblock {\em Electrochim. Acta}, \textbf{2000}, \textit{45}, 2559-2574.

\bibitem{Rickert871983}
H.~Rickert, H.-D.~Wiemh\"ofer,
\newblock {\em Ber. Bunsenges. Physikal. Chem.}, \textbf{1983}, \textit{87}, 236-239.

\end{thebibliography}

\newpage

\textbf{Supporting information}\\

\begin{figure}[ht]
  \begin{center}
\includegraphics[width=8cm]{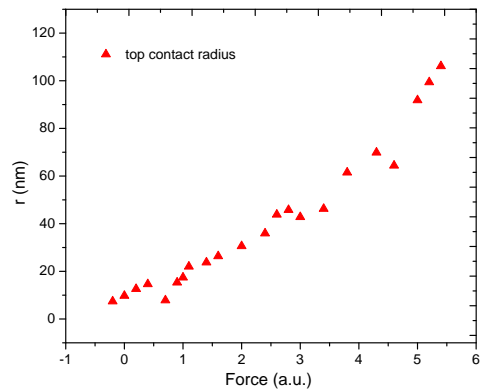}
\caption{Contact radius calculated from a set of experimental I-V curves using Eq.(\ref{I(V)}), at different load AFM tip forces. Measurements start when tip is just in contact with the surface of the Ag$_{2}$S (smallest contact radius) followed by an increase of the tip-sample interaction.}
\label{rvsF}
  \end{center}
\end{figure}  

\end{document}